\documentclass[
reprint,
showpacs,
preprintnumbers,
bibnotes,
amsmath,
amssymb,
aps,
prb,
floatfix,
longbibliography
]{revtex4-1}

\usepackage{graphicx}
\usepackage{dcolumn}
\usepackage{bm}
\usepackage{braket}
\usepackage{hyperref}

\usepackage{ulem}
\usepackage[usenames,dvipsnames]{color}

\begin{document}

\preprint{APS/123-QED}

\title{
Finite-temperature phase transition to a Kitaev spin liquid phase\\on a hyperoctagon lattice: A large-scale quantum Monte Carlo study
}

\author{Petr A. Mishchenko, Yasuyuki Kato, and Yukitoshi Motome}
\affiliation{Department of Applied Physics, University of Tokyo, Hongo 7-3-1, Bunkyo, Tokyo 113-8656, Japan}
\date{\today}

\begin{abstract}
The quantum spin liquid is an enigmatic quantum state in insulating magnets, in which conventional long-range order is suppressed by strong
quantum fluctuations. Recently, an unconventional phase transition was reported between the low-temperature quantum spin liquid and the
high-temperature paramagnet in the Kitaev model on a three-dimensional hyperhoneycomb lattice. Here, we show that a similar ``liquid-gas"
transition takes place in another three-dimensional lattice, the hyperoctagon lattice. We investigate the critical phenomena by adopting
the Green-function based Monte Carlo technique with the kernel polynomial method, which enables systematic analysis of up to $2048$ sites.
The critical temperature is lower than that in the hyperhoneycomb case, reflecting the smaller flux gap. We also discuss the transition on
the basis of an effective model in the anisotropic limit.
\end{abstract}

\pacs{Valid PACS appear here}

\maketitle
\section{Introduction}
The quantum spin liquid (QSL) is an exotic magnetic state of matter in solids, in which long-range ordering is suppressed by strong quantum
fluctuations even in the ground state~\cite{anderson_mater_res_bull_8_1973,fazekas_phil_mag_30_1974,balents_nature_464_2010}. As a
consequence of intensive study over several decades, substantial progress has been recently made in experimental search for
QSLs~\cite{kanoda_annu_rev_condens_matter_phys_2_2011,gingras_rpp_77_2014,okamoto_prl_99_2007,coldea_prb_68_2003,hiroi_jpsj_70_2001,
shores_jacs_127_2005,okamoto_jpsj_78_2009,han_prl_113_2014}. In such experiments, however, QSLs are usually identified by the absence of
any phase transition to a long-range ordered state. In other words, it is supposed that the low-temperature ($T$) QSL is adiabatically
connected to the high-$T$ paramagnetic state. This common belief relies on the magnetic analog of liquid helium $3$ which remains as a
liquid down to the lowest $T$ due to strong quantum fluctuations~\cite{anderson_mater_res_bull_8_1973}. Nonetheless, in conventional
fluids, gas and liquid can be distinguished by a phase transition. Hence, a similar phase transition can also be expected in the case of
magnetic states of matter, namely, between the ``spin gas" (paramagnet) and ``spin liquid" (QSL). However, the thermodynamics of QSLs
remains elusive, despite the crucial importance for understanding of existing and forthcoming experiments.
\par
Recently, an exotic finite-$T$ phase transition between the low-$T$ QSL and the high-$T$ paramagnet was reported for an extension of the
Kitaev
model defined on a three-dimensional (3D) hyperhoneycomb lattice as well as its anisotropic limit~\cite{nasu_prl_113_2014,
nasu_prb_89_2014}. The Kitaev model is an exactly-soluble quantum spin model, whose ground state provides a canonical example of
QSLs~\cite{kitaev_ann_phys_321_2006}, and has attracted attention as it might describe the magnetism in some spin-orbit entangled Mott
insulators\cite{khaliullin_prl_102_2009}. The finding of the exotic phase transition was brought by a newly-developed quantum Monte Carlo
(QMC) method based on a Majorana fermion representation, which is free from the negative sign problem. This phase transition is not
accompanied by any symmetry breaking, but can be explained by a proliferation of loops for excited $Z_2$ fluxes~\cite{kimchi_prb_90_2014,
nasu_prb_89_2014,nasu_prl_113_2014}. The emergent loop degree of freedom is specific to 3D; the finite-$T$ phase transition is absent and
turns into a crossover in two dimensions~\cite{nasu_prb_92_2015}. The surprising result urges reconsideration of both experimental and
theoretical quests for QSLs.
\par
As demonstrated in the hyperhoneycomb case~\cite{mandal_prb_79_2009}, the Kitaev model can be extended to any tri-coordinate
lattices~\cite{yang_prb_76_2007,brien_prb_93_2016,hermanns_prb_89_2014,yao_prl_99_2007}. Although all the extensions retain the QSL
nature~\cite{baskaran_prl_98_2007}, the exact ground state (the spatial configuration of $Z_2$ fluxes) is obtained for some limited cases
because Lieb's theorem is not applicable to generic tri-coordinate lattices~\cite{lieb_prl_73_1994}. Moreover, finite-$T$ properties have
not been studied in most cases. As there are a variety of 3D tri-coordinate lattices~\cite{brien_prb_93_2016,wells}, it is interesting to
investigate such 3D models for clarifying the universal aspects of the phase transition between the QSL and the paramagnet and for
exploring further exotic phase transitions.
\par
In this paper, we investigate finite-$T$ properties in the Kitaev model on another 3D lattice, the hyperoctagon
lattice~\cite{hermanns_prb_89_2014} (see Fig.~\ref{fig:fig_1}). For this model, the exact solution is not available, as
Lieb's theorem cannot be applied. Improving the QMC method by employing the Green function technique with the kernel polynomial method
(KPM), we compute much larger systems compared to the previous studies. From the systematic analysis of clusters up to $2048$ sites, we
find that the hyperoctagon model exhibits another example of finite-$T$ phase transitions between the QSL and the paramagnet. We find that
the critical temperature estimated from the careful analysis of the finite-size effects is lower than that for the hyperhoneycomb case,
reflecting the smaller flux gap~\cite{brien_prb_93_2016,kimchi_prb_90_2014}. In addition, we show that the anisotropic limit of the
hyperoctagon model becomes equivalent to that of the hyperhoneycomb one, which supports the common origin of the phase transition on two
lattices.
\par
The paper is structured as follows. In Sec.~\ref{sec:model_method}, we introduce the Kitaev model on a hyperoctagon lattice, and briefly
describe the numerical method used in the previous studies to analyze the finite-$T$ properties of the Kitaev model. Then we introduce the
improved QMC method by using the Green-function based KPM (GFb-KPM) and present the benchmark results. In Sec.~\ref{sec:results}, we
present numerical results. We identify the finite-$T$ phase transition and estimate the critical temperature from careful analyses of
finite-size effects for three different physical quantities. In Sec.~\ref{sec:discussion}, we discuss the origin of the phase transition
on the hyperoctagon lattice, with a consideration of the anisotropic limit of the model in comparison with that for the hyperhoneycomb
case. We also discuss a correlation between the critical temperature and the flux gap, with a brief comment on the ground state.
Finally, Sec.~\ref{sec:summary} is devoted to the summary.
\section{Model and Method\label{sec:model_method}}
\begin{figure}[htb]
\includegraphics[width=\columnwidth]{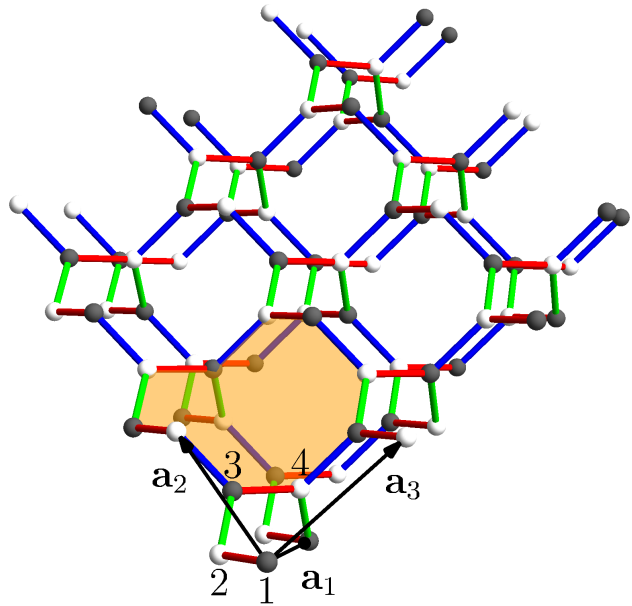}
\caption{\label{fig:fig_1}
Schematic picture of the hyperoctagon lattice. The spheres represent the lattice sites on which the spin-$1/2$ degrees of freedom are
defined in Eq.~(\ref{eq:kitaev_hamiltonian}). The sites $1$-$4$ represent the four-site unit cell, and ${\bf{a}}_\mu$ ($\mu = 1, 2, 3$)
represent the lattice translation vectors. The red, green, and blue bonds denote the $x$, $y$, and $z$ bonds in
Eq.~(\ref{eq:kitaev_hamiltonian}), respectively. The hyperoctagon lattice is bipartite, and the black $b$ and white $w$ spheres
identify the sublattices; see Eq.~(\ref{eq:majorana_hamiltonian}). The orange shade denotes an example of a ten-site plaquette on which
the $Z_2$ flux $W_p$ is defined.
}
\end{figure}

We study an extension of the Kitaev model~\cite{kitaev_ann_phys_321_2006} to a 3D hyperoctagon lattice shown in
Fig.~\ref{fig:fig_1}~\cite{hermanns_prb_89_2014}. The Hamiltonian is given by
\begin{equation}
\mathcal{H} = -J_x\sum_{\langle i,j \rangle_x}\sigma_i^x\sigma_j^x
               -J_y\sum_{\langle i,j \rangle_y}\sigma_i^y\sigma_j^y
               -J_z\sum_{\langle i,j \rangle_z}\sigma_i^z\sigma_j^z,
\label{eq:kitaev_hamiltonian}
\end{equation}
where $\sigma_i^x$, $\sigma_i^y$, and $\sigma_i^z$ are the Pauli matrices describing a spin-$1/2$ state at site $i$. The sum
$\langle i,j \rangle_\gamma$ is taken over the nearest-neighbor (NN) sites on the three different types of bonds, which are shown by red
($\gamma = x$), green ($y$), and blue ($z$) in Fig.~\ref{fig:fig_1}; $J_{\gamma}$ is the exchange constant for each bond. Similar to the
original Kitaev model on a honeycomb lattice~\cite{kitaev_ann_phys_321_2006}, the hyperoctagon model has a $Z_2$ conserved quantity called
the fluxes for each ten-site plaquette $p$ (see Fig.~\ref{fig:fig_1}),
$W_p = \prod_{\langle i,j \rangle_\gamma \in p} \sigma_i^\gamma \sigma_j^\gamma = \pm1$, where the product is taken for all the
bonds comprising the plaquette. While the ground state of this model is not exactly obtained, it was deduced to be a flux-free state,
where all $W_p$ are $+1$~\cite{hermanns_prb_89_2014}, similar to the honeycomb~\cite{kitaev_ann_phys_321_2006} and hyperhoneycomb
cases~\cite{mandal_prb_79_2009}.
\par
We investigate thermodynamic properties of the model in Eq.~(\ref{eq:kitaev_hamiltonian}) by the QMC simulation based on a Majorana
fermion representation~\cite{nasu_prl_113_2014}. In this technique, the lattice is regarded as an assembly of one-dimensional chains
composed of two types of bonds and the Jordan-Wigner transformation is applied to each chain~\cite{chen_prb_76_2007,feng_prl_98_2007,
chen_j_phys_a_41_2008}. By taking such chains along the ${\bf{a}}_1$ direction composed of the red and green bonds as in
Fig.~\ref{fig:fig_1}, the Kitaev Hamiltonian in Eq.~(\ref{eq:kitaev_hamiltonian}) is rewritten by two types of Majorana fermions $c$
and $\bar{c}$ as
\begin{equation}
\mathcal{H} = iJ_x\sum_{(w,b)_x}c_wc_b + iJ_y\sum_{(w,b)_y}c_wc_b + iJ_z\sum_{(w,b)_z}\eta_rc_wc_b, \label{eq:majorana_hamiltonian}
\end{equation}
where $\eta_r = i\bar{c}_b\bar{c}_w$ is a local conserved quantity taking $\pm 1$ ($r$ is the index for the $z$ bond); it satisfies
$W_p = \prod_{r\in p} \eta_r$, where the product is taken for all the $z$ bonds included in the ten-site loop. The sum of $(w,b)_\gamma$
is taken for the NN sites on the $\gamma$ bonds with $w > b$ ($b > w$) if the chain begins with $b$ ($w$) (see Fig.~\ref{fig:fig_1}). From
the form of Eq.~(\ref{eq:majorana_hamiltonian}), the system is regarded as free Majorana fermions coupled
with the $Z_2$ variables $\eta_r$. Hence, one can compute thermodynamic properties by Monte Carlo (MC) sampling over the configurations
$\{\eta_r\}$ with the statistical weight $P(\{\eta_r \}) = \exp[-\beta F(\{\eta_r \})]$, where $F(\{\eta_r \})$ is the free energy of
the Majorana fermions $\{c\}$ for a configuration $\{\eta_r\}$, and $\beta = 1/T$ is the inverse temperature (we set the Boltzmann
constant $k_{\rm B} = 1$). As $P(\{\eta_r\})$ is positive definite, the QMC simulation is free from the negative sign
problem~\cite{nasu_prl_113_2014,nasu_prl_115_2015, nasu_prb_92_2015,nasu_nphys_12_2016}.
\par
In the previous studies~\cite{nasu_prl_113_2014,nasu_prl_115_2015,nasu_prb_92_2015,nasu_nphys_12_2016}, $F(\{\eta_r\})$ was calculated by
the exact diagonalization (ED), which leads to the computational cost of $\mathcal{O}(N^4)$ for one MC sweep where $N$ is the number of
spins. This has limited the reachable system size $N$ to less than $10^3$ in a reasonable computational time. In order to reduce the
computational cost and enable systematic analysis of up to larger $N$, in the present study, we apply the GFb-KPM to estimate
$F(\{\eta_r \})$~\cite{weisse_prl_102_2009}. In this method, a change of the Majorana fermion density of states in a MC update is directly
computed from a small number of Green functions obtained by the KPM~\cite{weisse_rev_mod_phys_78_2006}, which reduces the computational
cost to $\mathcal{O}(N^2)$.
\par
Let us describe the method briefly. Suppose if the one-body Majorana Hamiltonian for a given configuration $\{\eta_r\}$,
$\mathcal{H}(\{\eta_r\})$, is modified to $\mathcal{H}(\{\eta_r\}) + \Delta$ by a local flip of $\eta_r$, then the energy spectrum is given
by the solution of $\mathcal{D}(E) = \mathrm{det}\{\mathbb{I} + G(E)\Delta\} = 0$, where $G(E)$ is the Green function satisfying
$G(E)\{\mathcal{H}(\{\eta_r\}) - E\mathbb{I}\} = \mathbb{I}$ ($\mathbb{I}$ is the unit matrix). By extending $\mathcal{D}(E)$ to a
complex function by $E \rightarrow E + i\epsilon$, the difference in the free energy by the local flip is given by
\begin{align}
&F(\{\eta_r'\}) - F(\{\eta_r\}) \nonumber \\
&= -\frac{N}{\pi}\int_{0}^{\infty}\lim_{\epsilon \to 0}
\operatorname{Im}\ln \{\mathcal{D}(E + i\epsilon)\} \frac{1}{2} \tanh\left(\frac{\beta E}{2}\right)dE.
\label{eq:freeenergy_change}
\end{align}
$\mathcal{D}(E + i\epsilon)$ can be calculated in a compact manner by using a few components of Green's
functions~\cite{weisse_rev_mod_phys_78_2006}. The Green functions are obtained by using the KPM with Chebyshev polynomials. For instance,
the onsite component is obtained as
\begin{equation}
G_{j,j}(E + i\epsilon) = i
\frac{{\mu_0} + 2\sum_{m = 1}^{M - 1}{\mu_m}\mathrm{exp}\{-im~\mathrm{arccos}(E/s)\}}
{\sqrt{s^2 - E^2}},
\label{eq:green_function_from_moments}
\end{equation}
where $\mu_m$ is the $j$th diagonal element of the $m$th Chebyshev moment of $\mathcal{H}(\{\eta_r \})$ rescaled by a factor of $s$ to fit
the eigenvalues in the range of $[-1:1]$, and $M$ is the cutoff of the expansion~\cite{weisse_prl_102_2009}. $\mu_m$ is calculated
efficiently by sparse matrix-vector multiplications~\cite{weisse_rev_mod_phys_78_2006}.
\par
In the following calculations, we focus on the isotropic case and set the energy unit as $J_x = J_y = J_z = 1/3$. We consider the clusters
from $32$ ($2^3$ unit cells) to $2048$ sites ($8^3$ unit cells) with the open boundary condition in the ${\bf{a}}_1$ direction and the
periodic ones in the remaining two directions. We typically perform ten independent MC runs with $500$ MC steps for measurements after
$3500$ to $5500$ steps for thermalization. In addition to MC sweeps by a single flip of $\eta_r$, we adopt the replica exchange MC
technique~\cite{hukushima_jpsj_65_1996}.
\par
\begin{figure}[htb]
\includegraphics[width=\columnwidth]{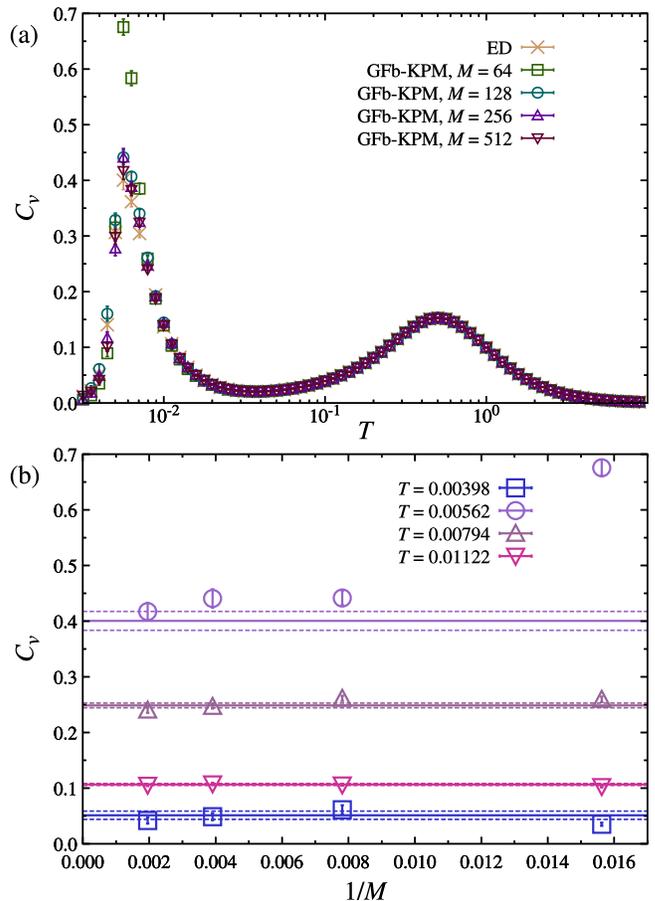}
\caption{\label{fig:fig_2}
(a) Comparison of the QMC results for the $T$ dependence of the specific heat per site, $C_v$, obtained by the update scheme with the ED
and GFb-KPM. In the GFb-KPM calculations, the Green functions are obtained by the Chebyshev expansion up to $M = 512$. (b) $M$ dependence
of $C_v$ at several temperatures in the low-$T$ region around the low-$T$ peak in (a). The horizontal solid lines represent the ED results
with errorbars indicated by the dashed lines. The calculations are done for the isotropic case $J_x = J_y = J_z = 1/3$ in
Eq.~(\ref{eq:kitaev_hamiltonian}) on the $4\times4^3 = 256$-site cluster.
}
\end{figure}

Figure~\ref{fig:fig_2} shows the benchmark of the GFb-KPM method. We compare the QMC data of the specific heat per site, $C_v$, computed
by the ED method and the GFb-KPM method for the $4\times4^3 = 256$-site cluster. For the GFb-KPM calculations, we present the data with a
different cutoff of the Chebyshev expansion, $M$, from $M = 64$ to $512$. Figure~\ref{fig:fig_2}(a) shows $T$ dependence of $C_v$. We find
that the GFb-KPM data well converge to the ED ones, even for the smallest $M = 64$, except for the low-$T$ region around the low-$T$ peak
of $C_v$. Figure~\ref{fig:fig_2}(b) shows the convergence of the GFb-KPM data in terms of $M$ in the low-$T$ region: the symbols indicate
the GFb-KPM data for each $M$, while the horizontal lines the ED ones (the dashed lines indicate the statistical errorbars). We find that
the GFb-KPM results quickly converge to the ED ones while increasing $M$. On the basis of this benchmark as well as the fact that the
GFb-KPM converges faster for larger system sizes~\cite{weisse_rev_mod_phys_78_2006}, we take $M = 512$ for the $576$-site cluster and
$M = 256$ for larger clusters in the following GFb-KPM calculations~\footnote{We compute the two smallest clusters with $N = 32$ and $256$
by QMC with ED.}.
\section{Results\label{sec:results}}
\begin{figure}[htb]
\includegraphics[width=\columnwidth]{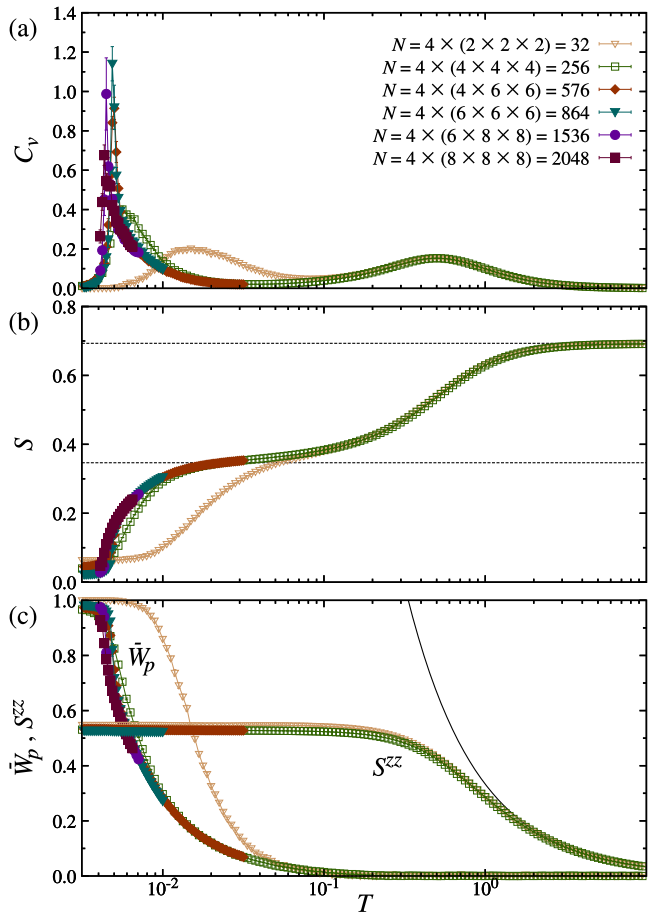}
\caption{\label{fig:fig_3}
Temperature dependence of (a) $C_v$, (b) the entropy per site $S$, and (c) the average of the $Z_2$ variables $W_p$, $\bar{W}_p$, and the
NN spin correlation on the $z$ bonds, $S^{zz}$. The calculations were performed for the isotropic case in Eq.~(\ref{eq:kitaev_hamiltonian})
on several clusters whose sizes are indicated in the figure. The dashed horizontal lines in (b) indicate $(1/2)\ln 2$ and $\ln 2$. The
black curve in (c) represents the high-$T$ asymptotic Curie behavior $1/(3T)$.
}
\end{figure}

Figure~\ref{fig:fig_3}(a) shows the QMC results of the specific heat per site, $C_v$. Similar to the hyperhoneycomb
case~\cite{nasu_prl_113_2014}, $C_v$ exhibits two peaks. The higher-$T$ peak is closely related with the development of NN spin
correlations $S^{zz} = 2 \sum_{\langle i,j \rangle_z} \langle \sigma_i^z\sigma_j^z \rangle / N$ shown in Fig.~\ref{fig:fig_3}(c),
which corresponds to the lowering of the kinetic energy of the itinerant Majorana fermions~\cite{nasu_prb_92_2015,nasu_prl_113_2014}. On
the other hand, the lower-$T$ peak in $C_v$ is associated with the coherent alignment of the $Z_2$ variables $W_p$~\cite{nasu_prb_92_2015,
nasu_prl_113_2014}; see the thermal average $\bar{W}_p = \langle \sum_p W_p \rangle / N_p$ plotted in Fig.~\ref{fig:fig_3}(c) ($N_p$ is
the number of ten-site plaquettes in the system). Figure~\ref{fig:fig_3}(b) shows the results of the entropy per site,
$S = \ln 2 + \beta\langle\mathcal{H}\rangle/N - \int_0^\beta \langle\mathcal{H}\rangle d\beta/N$. At each peak of $C_v$, a half of
$\ln 2$ entropy is released; the higher-$T$ release comes from itinerant Majorana fermions and the lower one comes from $Z_2$ variables
$W_p$. Similar behavior was reported for the Kitaev models on the honeycomb and hyperhoneycomb lattices~\cite{nasu_prb_92_2015,
nasu_prl_113_2014}. This is the common feature in the Kitaev QSLs called thermal fractionalization of quantum
spins~\cite{nasu_prb_92_2015}.
\par
\begin{figure}[htb]
\includegraphics[width=\columnwidth]{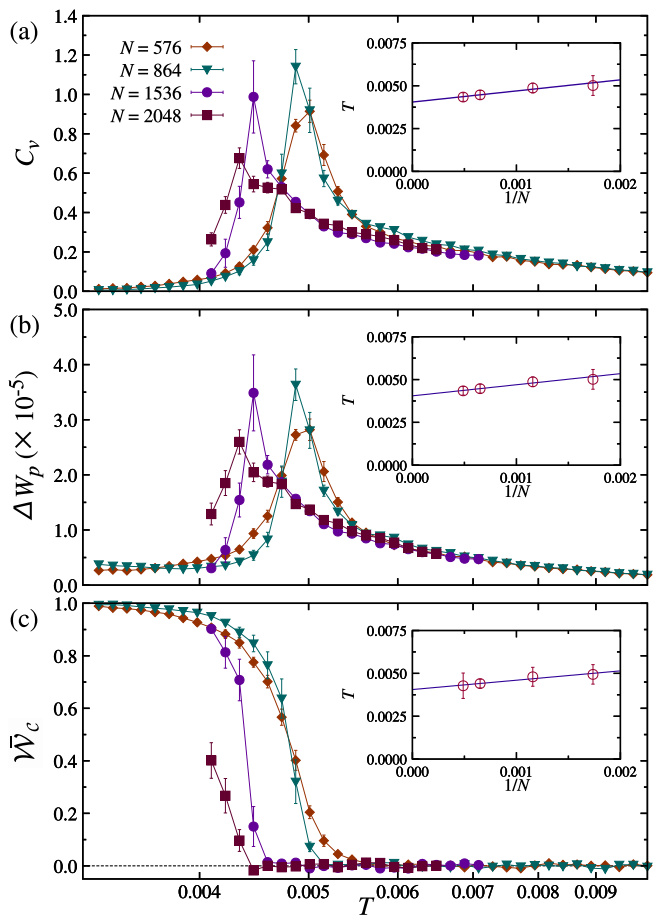}
\caption{\label{fig:fig_4}
Enlarged views of the low-$T$ region: (a) $C_v$, (b) fluctuations of $W_p$, $\Delta W_p$, and (c) the Wilson loop $\bar{\mathcal{W}}_c$.
The insets show the system-size extrapolation of the peak $T$ of (a) $C_v$ and (b) $\Delta W_p$, and (c) the inflection point of $T$
dependence of $\bar{\mathcal{W}}_c$.
}
\end{figure}

In Fig.~\ref{fig:fig_3}(a), while the high-$T$ peak of $C_v$ is almost system-size independent, the low-$T$ peak exhibits significant
system-size dependence: the peak becomes sharper and shifts to lower $T$ while increasing the system size. The enlarged view of the low-$T$
part is shown in Fig.~\ref{fig:fig_4}(a). The system-size dependence is indicative of a phase transition, as in the hyperhoneycomb
case~\cite{nasu_prl_113_2014}. We note that the height of the peaks does not show systematic behavior, which is presumably due to the
different shapes of clusters. We find similar critical behavior in thermal fluctuations of $W_p$, defined by
\begin{align}
\Delta W_p = \frac{(\sum_\gamma J_\gamma)^2}{N_p T^2}
\Big[ \big\langle \big( \sum_p W_p \big)^2 \big\rangle - \big\langle \sum_p W_p \big\rangle^2 \Big].
\end{align}
As shown in Fig.~\ref{fig:fig_4}(b), $\Delta W_p$ shows a similar peak to $C_v$. We note that this quantity gives a measure of the
specific heat in the case of the anisotropic limit of the Kitaev model where the Hamiltonian is described by $W_p$ only; namely,
$\Delta W_p$ gives a measure of energy fluctuations related to $W_p$. Thus, the common critical behavior between $C_v$ and $\Delta W_p$
indicates that the phase transition is driven by the $Z_2$ variables $W_p$.
\par
In order to characterize the phase transition, following the previous study for the hyperhoneycomb case~\cite{nasu_prl_113_2014}, we also
measure the Wilson loop $\bar{\mathcal{W}}_c$ on a plane perpendicular to one of the periodic boundaries\footnote{We define
$\bar{\mathcal{W}}_c$ on a half of the largest loop on the plane which is halved in the periodic boundary condition.}. As shown in
Fig.~\ref{fig:fig_4}(c), $\bar{\mathcal{W}}_c$ is zero at high $T$ and grows rapidly in well accordance with the peaks of $C_v$ and
$\Delta W_p$. This behavior is in contrast to the gradual growth of $\bar{W}_p$ in Fig.~\ref{fig:fig_3}(c). The result indicates that the
global quantity $\bar{\mathcal{W}}_c$ acts as an order parameter for the unconventional phase transition, as in the hyperhoneycomb
case~\cite{nasu_prl_113_2014}.
\par
From the critical behaviors of $C_v$, $\Delta W_p$, and $\bar{\mathcal{W}}_c$, we estimate the critical temperature $T_c$. The insets in
Fig.~\ref{fig:fig_4} present the system-size extrapolation of the peak $T$ of $C_v$ and $\Delta W_p$ and the inflection point of $T$
dependence of $\bar{\mathcal{W}}_c$. The extrapolated values to $N \to \infty$ well coincide with each other: $T_c = 0.00405(10)$ by
$C_v$, $0.00405(10)$ by $\Delta W_p$, and $0.00406(8)$ by $\bar{\mathcal{W}}_c$ (the numbers in the parentheses represent the errors in
the last digits). This good agreement indicates the validity of our numerical simulation and the system-size analysis.
\section{Discussion\label{sec:discussion}}
From the comparison to the previous study for the hyperhoneycomb model~\cite{nasu_prl_113_2014}, we conclude that the phase transition on
the hyperoctagon lattice is another realization of the ``liquid-gas" phase transition between the low-$T$ QSL and the high-$T$ paramagnet.
Similar to the hyperhoneycomb case, we do not observe any indication of discontinuity in the transition~\footnote{We carefully examine the
histogram of the internal energy, but do not find any clear signature of bifurcation.}. In the hyperhoneycomb case, the origin of the phase
transition is ascribed to a proliferation of closed loops formed by flipped $W_p$~\cite{nasu_prl_113_2014}. Such loop excitations are
commonly seen in the 3D systems including the hyperoctagon case~\cite{kimchi_prb_90_2014,hermanns_prb_89_2014}. Our result suggests that
the phase transition takes place universally in the 3D Kitaev models once the low-energy physics is described by closed loops of $W_p$.
\par
The common phase transition between the hyperoctagon and hyperhoneycomb cases is also inferred by considering the anisotropic limit of the
models, $J_z \gg J_x$ and $J_y$. For both cases, the low-energy effective model is obtained by the similar perturbation theory to that
used for the derivation of the toric code from the Kitaev model in Ref.~\onlinecite{kitaev_ann_phys_321_2006}. In the perturbation
theory, the original Hilbert space is projected onto the low-energy subspace where each dimer made of two spins connected by the strong $z$
bond, say $\sigma_1$ and $\sigma_2$, takes only two states out of four; these two states are represented by a pseudospin
$| \tau^z_{ij} = \pm 1 \rangle = | \sigma^z_i = \sigma^z_j = \pm 1 \rangle$. When one defines the pseudospins at the centers of $z$
bonds, the lattice is reduced to the diamond lattice for both hyperoctagon and hyperhoneycomb cases. For the hyperhoneycomb case, the
low-energy effective model was obtained in Ref.~\onlinecite{mandal_prb_79_2009} by considering the eighth-order perturbation. The
effective model is described by the projected $Z_2$ fluxes $B_p =\pm 1$ that are defined for each six-site plaquette in the diamond
lattice, and has the form of $\mathcal{H}_{\rm eff} = -\sum_p  J^p_{\rm eff} B_p$, where $J^p_{\rm eff}$ is either $7J_x^4J_y^2/256J_z^5$
or $7J_y^4J_x^2/256J_z^5$ depending on the number of $x$ bond included in the plaquette $p$. It was numerically demonstrated that the
effective model exhibits a finite-$T$ phase transition between the high-$T$ paramagnet and the low-$T$ QSL~\cite{nasu_prb_89_2014}.
Performing the similar procedure, we find that the low-energy effective model for the hyperoctagon case has the same form, except for the
sign of the coupling constant $J^p_{\rm eff}$. As the opposite sign does not cause any difference in the thermodynamics, the effective
model shows the same finite-$T$ phase transition as in the hyperhoneycomb case. Although this correspondence is only in the anisotropic
limit, it strongly suggests a common mechanism of the finite-$T$ phase transition between the two models even in the isotropic case.
\par
On the other hand, we note that our estimate of $T_c$ for the hyperoctagon model is lower than that for the hyperhoneycomb case, as shown
in Table~\ref{table:table_0}~\cite{nasu_prl_113_2014}. As the phase transition is caused by the loop proliferation, $T_c$ is closely
related with the loop tension~\cite{kimchi_prb_90_2014}. The loop tension is determined by the flux gap $\Delta$, which is defined by the
minimum excitation energy for flipping $W_p$ from the ground state. The flux gap is estimated as $\Delta = 0.030(3)$ for the hyperoctagon
lattice and $\Delta = 0.043(3)$ for the hyperhoneycomb lattice~\cite{brien_prb_93_2016}. The values of $T_c$ and $\Delta$ summarized in
Table~\ref{table:table_0} indicate a good correlation between $T_c$ and $\Delta$.
\par
Finally, let us briefly comment on the ground state. As shown in Fig.~\ref{fig:fig_3}(c), our QMC data for $\bar{W}_p$ converge to $+1$ at
low $T$. This suggests that the system is likely to be in the flux-free state with all $W_p = +1$ when $T \to 0$, as deduced by variational
arguments in the previous study~\cite{hermanns_prb_89_2014}.
\begin{table}[htb]
\begin{center}
\begin{tabular}{|c|c|c|} \hline
Lattice        & Critical temperature $T_c$ & Flux gap $\Delta$ \\ \hline \hline
Hyperoctagon   & 0.00405(10)                & 0.030(3)            \\ \hline
Hyperhoneycomb & 0.00519(9)                 & 0.043(3)            \\ \hline
\end{tabular}
\caption{\label{table:table_0}
Comparison between the critical temperature $T_c$ obtained by the QMC simulations and the magnitude of the flux gap $\Delta$ estimated at
$T = 0$ for the hyperhoneycomb and hyperoctagon cases. $T_c$ for the hyperhoneycomb system is taken from
Ref.~\onlinecite{nasu_prl_113_2014}, while $\Delta$ from Ref.~\onlinecite{brien_prb_93_2016}.
}
\end{center}
\end{table}
\section{Summary\label{sec:summary}}
In summary, we have investigated the ``liquid-gas" phase transition from the low-$T$ QSL to the high-$T$ paramagnet on the extension of the
Kitaev model to the 3D hyperoctagon lattice. Using an improved QMC method by the Green function technique with the KPM, we successfully
dealt with much larger-size systems compared to the previous studies. From the systematic analysis of clusters up to $2048$ sites, we found
a finite-$T$ phase transition similar to the previous hyperhoneycomb case. The result was also supported by considering the anisotropic
limit of the model on two lattices. We also showed that the critical temperature on the hyperoctagon lattice is lower than in the
hyperhoneycomb case, reflecting the smaller flux gap in the former case. It will be interesting to extend the present study to other 3D
Kitaev models for an exploration of further exotic phase transitions.
\begin{acknowledgments}
The authors thank J. Nasu, R. Ozawa, and J. Yoshitake for fruitful discussions. This research was supported by JSPS Grants-in-Aid for
Scientific Research Grants No.~JP15K13533, No.~JP16H02206, and No.~26800199. Parts of the numerical calculations were performed in the
supercomputing systems in ISSP, the University of Tokyo.
\end{acknowledgments}
\bibliography{prb}
\end{document}